\documentstyle[12pt,aps]{revtex}

\begin{document}
\title{
\hfill\parbox[t]{2in}{\rm\small\baselineskip 14pt
{~~~~~JLAB-THY-99-09}\vfill~}\\
\hfill\parbox[t]{2in}{\rm\small\baselineskip 14pt
{~~~~~~~~~~~~~~(version 2)}\vfill~}
\vskip 2cm
Flux Tube Zero-Point Motion, Hadronic Charge Radii, and 
Hybrid Meson Production Cross Sections}

\vskip 1.0cm

\author{Nathan Isgur}
\address{Jefferson Lab, 12000 Jefferson Avenue,
Newport News, Virginia, 23606}
\maketitle

\vspace{2.0 cm}
\begin{center}  {\bf Abstract}  \end{center}
\vspace{.4 cm}
\begin{abstract}

	Flux tube zero-point motion produces quark displacements transverse  to 
the flux tube which make
significant contributions to hadronic charge radii.  In heavy quark systems, these
contributions can be related by Bjorken's
sum rule to the rates for semileptonic decay to hybrid mesons.  This connection
can be generalized to other
leptoproduction processes, where transverse contributions to
elastic form factor slopes are related to the cross sections for the production of the associated
hybrid states.  I identify the flux tube overlap integral responsible for these effects
as the strong QCD analogue of the Sudakov form factor of perturbative QCD.

\bigskip\bigskip\bigskip\bigskip

\end{abstract}
\newpage

\section {Introduction}
\medskip
		The discovery and study of hybrid mesons is one of the most urgent tasks in
strong interaction physics today, since their properties are a direct reflection
of the mechanism of confinement. Theory \cite{IP,ConfinementIII} has provided compelling arguments from QCD
that confinement occurs {\it via} the formation of a flux tube.  In the simplest situation,
corresponding to a long tube with fixed $Q$ and $\bar Q$ sources on its ends, a flux 
tube has a very simple vibrational spectrum corresponding to the excitation of 
transverse phonons in its string-like structure.  The essential features of this
gluonic spectrum are retained in the spectrum of real mesons with their flux tube
excited (the hybrids), and consequently searches for hybrid mesons (especially
those with tell-tale $J^{PC}$ exotic quantum numbers) are underway in many 
laboratories \cite{BNLhybrid,otherhybrid}. 

	These searches will probably be technically difficult since the dominant final states resulting
from the decays of the lightest hybrids are predicted \cite{IKP,ClosePage} to be very complex  multiparticle
states arising from the decay of intermediate  $S + P$ decay products
(where $S$ is an $\ell=0$ ground state meson and $P$  an $\ell=1$ excited state).  
In addition to their technical difficulty, hybrid meson searches have been considered
difficult to interpret because hybrid meson production cross sections had only been 
estimated on very qualitative grounds \cite{IKP}. The qualitative argument
was frighteningly simple:  a flux tube is a strongly interacting
object, so it should be expected to be excited as readily as a quark.  It followed that exclusive
hybrid meson production cross sections should be comparable to exclusive ordinary
meson production cross sections.  This argument has recently been supported by a
detailed Regge-theory-based calculation \cite{AfanasevPage} for the electro- and photoproduction of the 
$J^{PC} = 1^{-+}$ hybrid. 

	  Using Bjorken's heavy quark sum rule \cite{Bj,IWonBj}, I present here an argument which
relates the inclusive production cross section of hybrid mesons in heavy quark decay to a new
contribution to the slope $\rho^2$ of the Isgur-Wise function \cite{IWoriginal}.  
This new contribution to elastic form factors is remarkably simple in character, but 
I can find no reference to it in the existing literature:  zero-point
oscillations of the confining flux tube will lead to transverse displacements of a meson's
quarks from the interquark axis.  These displacements contribute to the charge radius
of a quark, making the elastic form factor fall faster than it would have otherwise.  
By Bjorken's sum rule (and its generalizations),
this loss of rate from the elastic channel must be compensated by an inelastic channel.  
At the mechanical level, the loss of rate to the elastic channel occurs because the 
ground state wavefunctions of the flux tubes in the initial and final meson states do
not overlap perfectly.  It is then intuitively clear that the lost rate will appear in
final states where the flux tube is not in its ground state, {\it i.e.}, hybrid mesons.  In
what follows I will demonstrate this explicitly and show how the inclusive hybrid
production cross sections are distributed into exclusive modes.

This connection between charge radii and hybrid production rates shows that the 
qualitative argument for the strength of hybrid meson
cross sections presented in Ref. \cite{IKP}
was correct:  hybrid production proceeds without suppression by any small parameters (nor 
any factors of $1 / N_c$).  Bjorken's sum rule is only exact as a local duality in the 
heavy quark limit, but it can be generalized \cite{CabibboRadicati} via the operator
product expansion (OPE) to an approximate global duality between hybrid meson and baryon
electroproduction and a new contribution to the elastic charge radii of light hadrons.

	 \medskip
\section {Transverse Contributions to Hadronic Charge Radii}
\medskip
	In nonrelativistic quark potential models for $Q\bar d$ mesons, each quark develops a charge
radius by making longitudinal excursions from the quark-antiquark center of mass along
the interquark axis $\vec r \equiv \vec r_{\bar d} - \vec r_Q$.  (I have employed a notation 
here that is suggestive of a heavy quark meson where $m_Q \gg m_d$ since Bjorken's sum rule
applies exactly to this case, but I will keep open the possibility of arbitrary masses in most
of the following.)  In the nonrelativistic, valence approximation, $\vec r_Q = -({m_d \over {m_Q + m_d}}) \vec r$
so the squared charge radius of $Q$ is  
\begin{equation}
r^2_Q=\Bigl( {m_d \over {m_Q+m_d}}\Bigr)^2 \langle r^2 \rangle
\label{eq:nrchargeradius}
\end{equation}
where $\langle r^2 \rangle$ is the expectation value of the squared interquark separation
in the meson wavefunction $\psi_{Q \bar d} (\vec r)$.  If the hadron in question has additional
nonrelativistic constituents, as does a baryon, this relation is 
trivially modified with $\vec r$ interpreted as
the relative coordinate of the center of mass of the constituents accompanying $Q$.

	The flux tube between $Q$ and $\bar d$ is a relativistic object with an infinite number of degrees
of freedom.  For this reason the flux tube model is usually simplified by employing an adiabatic
approximation which converts it into a quark potential model \cite{IP}.  
This approximation consists
in first ``nailing down" the ends of the flux tube and solving for its quantum states in the limit
where it behaves like a thin piece of relativistic string with its purely transverse degrees of
freedom.  The resulting energies $E_n (r)$ ($n$ defines the string's quantum state) are then
used as (adiabatic or Born-Oppenheimer) effective potentials $V_n(r)$ on which meson 
spectroscopies are built:  ordinary mesons are vibrational or rotational states built on the
gluonic ground state effective potential $V_0(r)$, the lightest hybrids are built on the 
``one-phonon with wavelength $2r$" potential $V_1 (r)$, {\it etc.}  In this approximation, as in any nonrelativistic potential
model, Eq. (\ref{eq:nrchargeradius}) continues to hold.  However, 
for any finite quark mass, transverse fluctuations of the 
flux tube will lead to transverse excursions of $Q$ and a correction to this equation.  In the next
Section I derive the result
\begin{equation}
r^2_Q=\Bigl[ \Bigl( {m_d \over {m_Q+m_d}}\Bigr)^2 + {2b \over \pi^3 m^2_Q} \zeta(3) \Bigr]\langle r^2 \rangle
\label{eq:r2correction}
\end{equation}
or, in the heavy quark limit
\begin{equation}
\rho^2={1 \over 3} \Bigl[ m_d^2 + {2b \over \pi^3 } \zeta(3) \Bigr]\langle r^2 \rangle~~,
\label{eq:rho2correction}
\end{equation}
where $\zeta(3) \equiv \sum_{p=1}^\infty {1 \over p^3} \simeq 1.20$.
Note that, like the potential model term, the correction terms from 
flux-tube transverse oscillations are
proportional to the mean square interquark distance.  This occurs because the mass 
of the vibrating flux tube,
to which $r^2_{Q\perp}$ is proportional, is $br$.  Using ``canonical parameters" \cite{IP,ISGW2},
Eq. (\ref{eq:r2correction}) 
gives 
a huge $51 \%$ correction 
in light quark systems where $m_Q = m_d$; in heavy quark systems Eq. (\ref{eq:rho2correction}) gives a $13 \%$ 
correction.  In
light quark systems this missing transverse contribution could, along with now
well-established relativistic corrections \cite{relcorrections}, be responsible for 
the infamous 
underprediction of charge radii in quark models.  

	  The precise numerical values I have quoted for these corrections should not be taken too seriously, since the gluonic
fields in ground state mesons are certainly not pure flux tubes.  However, lattice QCD 
gives the remarkable result that
$V_0 (r)$ becomes linear at such small distances that much of even ground state wavefunctions
may be flux tube dominated.  This simple fact can
only suggest that the qualitative
character of the correction terms in Eqs. (\ref{eq:r2correction}) and (\ref{eq:rho2correction})
will not be lost.  However, I believe there is a compelling general
physics case for such strong transverse corrections.  Consider the $Q \bar d$ system
immediately after the $Q$ makes a sudden transition from rest (as $m_Q \rightarrow \infty$, it has
$\vec v = 0$) to velocity $\vec v~'$.  Independent of their geometry, 
by Gauss' law the color electric field lines in the neighborhood of $Q$ must
immediately begin to follow it, but by causality those in the neighborhood of $\bar d$ must continue to point
in the original direction of $Q$ for a time $T \geq r/c$.
A ``kink" will therefore be formed in the classical color electric field lines which will propagate from
$Q$ to $ \bar d$ with the period $T$.  This classical kink in the color electric field will in turn
manifest itself quantum mechanically in the excitation of gluonic degrees of freedom.  (The preceeding
description is the QCD analog of the textbook example 
presented in Ref. \cite{Purcell} for classical electromagnetism where the analogous excitation results
in classical radiation and thus to the quantum emission of photons.  In perturbative QCD 
the acceleration of $Q$ would result in the emission of gluons;
in the strong QCD \cite{CloseQuote} regime  of interest here the excitation is confined, making a hybrid meson.)
Thus while flux tube dynamics provides a concrete model in which these transverse excitations
and the associated transverse contributions to charge radii required by Bjorken's sum rule can be realized,
it seems likely that the basic phenomenon is very general. In particular, even 
a Coulomb-like gluonic field would lead to similar effects of comparable strength
(since $\alpha_s \sim 1$ in these systems).

	As a prelude to the next Section, I note that the sum $\zeta(3) \equiv \sum {1 \over p^3}$ in
Eq. (\ref{eq:rho2correction}) is $\sim 80 \%$ saturated by its $p =1$ term.
This strongly suggests what we will discover by explicit calculation below:  hybrid
production is dominantly in the lowest phonon mode.  We will also see that a substantial fraction of this
production goes into the mesonic ground state of the associated adiabatic surface $V_1 (r)$.     
\medskip
\section {Hybrid Excitation and Bjorken's Sum Rule}
\medskip

  To determine the contribution to $ \langle {\vec  r_Q}^{~2} \rangle$
of the motion of the flux tube or to calculate the transition rates to the states in which the flux 
tube is excited, it is  necessary to explicitly introduce the flux tube degrees of freedom
into the description of hadrons.
To do this, I adopt the method of Appendix A of Ref. \cite{KI} in which the flux tube is discretized into $N$ cells.
In the adiabatic approximation in which the points $n=0$ and $N+1$ are fixed on static quarks and 
$n=1$, $2$, $...N$ have transverse degrees of freedom $\vec y_n$, the state of the flux tube can be written
in terms of a complete set of transverse eigenstates
\begin{equation}
\vert \vec y \rangle = \vert \vec y_1  \vec y_2 \cdot \cdot \cdot   \vec y_N   \rangle . 
\end{equation}
Alternatively, and more conveniently, the Fourier coefficients
\begin{equation}
\vec a_p \equiv \sum_{n=1}^N \vec y_n \sqrt{2 \over {N+1}} sin {p \pi \over {N+1}}n
\end{equation}
can be introduced and the state of the flux tube can be written in terms of a complete set of Fourier
amplitudes
\begin{equation}
\vert \vec a \rangle = \vert \vec a_1  \vec a_2 \cdot \cdot \cdot   \vec a_N   \rangle . 
\end{equation}
The classical Lagrangian for this system in the small oscillations approximation is
\begin{equation}
L(\vec y,{d\vec y \over dt})=ba \sum_{n=0}^N \Bigl[ {1 \over 2} ({d\vec y_n \over dt})^2 
- {1 \over 2a^2} (\vec y_{n+1}-\vec y_n)^2\Bigr]
\end{equation}
where $b$ is the string tension and $a$ is the length of each cell.  Transforming to the Fourier
coefficients gives
\begin{equation}
L(\vec a,{d\vec a \over dt})=ba \sum_{p=1}^N \Bigl[ {1 \over 2} ({d\vec a_p \over dt})^2 
- {1 \over 2}\omega_p^2 \vec a_p^2 \Bigr]
\label{eq:aLagrangian}
\end{equation}
where
\begin{equation}
\omega_p \equiv {2 \over a}sin { \pi p\over 2(N+1)}~~.
\end{equation}
In the Fourier basis the eigenstates can be labelled by occupation numbers $n^{\alpha}_p$, the number of ``phonons" in 
mode $p$ with transverse polarization $\alpha=1$ or $2$:
\begin{equation}
\vert n \rangle = \vert n_1^1   n_1^2   n_2^1  n_2^2 \cdot \cdot \cdot    n_N^1  n_N^2   \rangle  
\end{equation}
with $E_n=\sum_{p=1}^N(n_p^1+n_p^2+1) \omega_p$ and a wavefunction
\begin{equation}
\chi_n(\vec a) \equiv \langle \vec a \vert n \rangle = 
\chi_{n_1}(a_1^1)  \chi_{n_1}(a_1^2)  \cdot \cdot \cdot \chi_{n_N}(a_N^1)  \chi_{n_N}(a_N^2) 
\end{equation}
where
\begin {equation}
\chi_{0_p} (a_p^{\alpha})={ {\alpha_p^{1/2}} \over {\pi^{1 / 4}}}
exp[-{1 \over 2}\alpha_p^2 (a_p^{\alpha})^2]
\end {equation}
and
\begin {equation}
\chi_{1_p}(a_p^{\alpha})=\sqrt{2} \alpha_p a_p^{\alpha} \psi_{0_p^{\alpha}}(a_p^{\alpha})
\end {equation}
are the wavefunctions we will need here.  Note that
\begin {equation}
\alpha_p^2=2b sin{  {\pi p} \over  {2(N+1)} }
\end {equation}
follows from Eq. (\ref{eq:aLagrangian}).

	To study the effects of the flux tube on form factors, we must go beyond the static approximation
to consider the normal modes of the flux tube-quark system.  For large quark masses $m_d$, 
$m_Q \gg br$, the transverse oscillations of frequency $\omega_p=\pi p/r$ 
are fast relative to the nearly static longitudinal motion, and so these
normal modes may be considered to be at fixed $r$. 
(In the low-lying states of a linear potential, the ratio of these two frequencies is
$(\mu_{Q \bar d}^2/b)^{1 \over 3}$, where $\mu_{Q \bar d}$ is the $Q \bar d$ reduced mass.)
In these circumstances one can easily show
using conservation of the position of the center of mass and of orbital angular momentum about
the center of mass that the mode $p$ with amplitude $\vec a_p$ involves transverse excursions
of the $Q$ and $\bar d$ quarks given by 
\begin {eqnarray}
\vec r_{Q_\perp} &=& -{br \over \pi p m_Q} \sqrt{2 \over {N+1}} \vec a_p 
\label{eq:rQperp} \\
\vec r_{\bar d_\perp} &=& (-1)^p{br \over \pi p m_d} \sqrt{2 \over {N+1}} \vec a_p ~~.
\label{eq:rdperp}
\end {eqnarray}

	We are now in a position to write down the full flux tube model state vectors of a $Q \bar d$ meson 
located at center of mass position $\vec R$.  The $n^{th}$ normal $Q \bar d$
meson $M_n^{(0)}$ built on the adiabatic potential $V_0 (r)$ corresponding to the flux tube 
ground state is
\begin {eqnarray}
\vert M_n^{(0)}(\vec R) \rangle &=& \int d^3r \int d^2a_1 \cdot \cdot \cdot \int d^2a_N 
\psi_n^{(0)}(\vec r)
\chi_{0_1}(a_1^1)
\chi_{0_1}(a_1^2)
\cdot \cdot \cdot  \nonumber \\
&&    
\chi_{0_N}(a_N^1)
\chi_{0_N}(a_N^2)\vert Q(\vec R-{ {m_d+{1 \over 2}br} \over {m_Q+m_d+br}} \vec r - {br \over \pi m_Q}\sqrt{2 \over {N+1}} 
\sum_p {1 \over p}\vec a_p);  \nonumber \\
&&
\vec a_1 \vec a_2 \cdot \cdot \cdot  \vec a_N ;
d(\vec R+{ {m_Q+{1 \over 2}br} \over {m_Q+m_d+br}} \vec r + {br \over \pi m_d}\sqrt{2 \over {N+1}} 
\sum_p {(-1)^p \over p}\vec a_p) \rangle
\end {eqnarray}
while the $n^{th}$ hybrid $Q \bar d$ meson built on the potential $V_1 (r)$ corresponding to the
excitation of one phonon in the lowest modes $p=1$, $\alpha=1$, and $p=1$, $\alpha=2$ are
\begin {eqnarray}
\vert H_{n\ell m}^{(1\pm)}(\vec R) \rangle &=& \int d^3r \int d^2a_1 \cdot \cdot \cdot \int d^2a_N 
\psi_{n\ell m}^{(1\pm)}(\vec r) \alpha_1(a_1^1 \pm i a_1^2)
\chi_{0_1}(a_1^1)
\chi_{0_1}(a_1^2)
\cdot \cdot \cdot   \nonumber \\
&&  
\chi_{0_N}(a_N^1)
\chi_{0_N}(a_N^2)
\vert Q(\vec R-{ {m_d+{1 \over 2}br} \over {m_Q+m_d+br}} \vec r - {br \over \pi m_Q}\sqrt{2 \over {N+1}} 
\sum_p {1 \over p}\vec a_p);  \nonumber \\
&&
\vec a_1 \vec a_2 \cdot \cdot \cdot   \vec a_N ;
d(\vec R+{ {m_Q+{1 \over 2}br} \over {m_Q+m_d+br}} \vec r + {br \over \pi m_d}\sqrt{2 \over {N+1}} 
\sum_p {(-1)^p \over p}\vec a_p) \rangle~~.
\end {eqnarray}
In these expressions the quarks $Q$ and $\bar d$ are in position eigenstates which include their 
equilibrium separation $\vec r$ and their excursions due to Eqs. (\ref{eq:rQperp}) and (\ref{eq:rdperp}), while the flux tube state is described in a
normal mode expansion with respect to its degrees of freedom $a^\alpha_p$ transverse to $\vec r$.
Good orbital angular momentum states in the hybrid sector have been formed by using 
circularly polarized phonon states
built out of $\alpha=1$ and $2$ with angular momenta $\pm \hbar$ around the axis 
$\vec r \leftrightarrow (r \theta \phi)$ so that, for example
\begin {eqnarray}
\psi_{n11}^{(1+)} &=& \sqrt{3 \over 4\pi}\tilde \psi_{H_n^{(1)}}(r)[{{1+cos \theta} \over 2}] 
\label{eq:psi11hybrid} \\
\psi_{n10}^{(1+)} &=& \sqrt{3 \over 4\pi}\tilde \psi_{H_n^{(1)}}(r)[\sqrt{1 \over 2}sin \theta] e^{-i\phi}
\label{eq:psi10hybrid} \\
\psi_{n1-1}^{(1+)} &=& \sqrt{3 \over 4\pi}\tilde \psi_{H_n^{(1)}}(r)[{{1-cos \theta} \over 2}] e^{-2i\phi} 
\label{eq:psi1-1hybrid}
\end {eqnarray}
where $\tilde \psi_{H_n^{(1)}}(r)$ is normalized to $\int drr^2 \vert \tilde \psi_{H_n^{(1)}}(r) \vert^2=1$.  
Note that the normal mode expansion is with respect to
body-fixed unit vectors which can be expanded as
\begin {eqnarray}
\hat e_1(\vec r) &=& [({{1+cos \theta} \over 2})-({{1-cos \theta} \over 2})cos 2\phi]~ \hat x
-({{1-cos \theta} \over 2})sin 2\phi~ \hat y
-sin \theta cos \phi~ \hat z  \\
\hat e_2(\vec r) &=& 
-({{1-cos \theta} \over 2})sin 2\phi~ \hat x
+[({{1+cos \theta} \over 2})+({{1-cos \theta} \over 2})cos 2\phi]~ \hat y
-sin \theta sin \phi~ \hat z 
\label{eq:transvectors} 
\end {eqnarray}
relative to the space-fixed coordinate axes $\hat x$, $\hat y$, $\hat z$.  
These expressions follow from a rigid rotation of the flux tube from $r \hat z$ to $\vec r$ in the $(r \hat z,\vec r)$ plane.
Finally, before proceeding to calculate form factors, we form total momentum eigenstates
\begin {equation}
\vert \vec P \rangle = \int d^3R {e^{i\vec P \cdot \vec R} \over (2 \pi)^{3 / 2}} \vert \vec R \rangle
\end {equation}
where $\vert \vec R \rangle$ is either set of position eigenstates.

	I begin by using this machinery to calculate the elastic form factor for scattering the 
pseudoscalar ground state $P \equiv M^{(0)}_0$ from momentum $-{\vec P_{cm} / 2}$ to $+ {\vec P_{cm} / 2}$ 
{\it via} the current $\bar Q \Gamma Q$.  Since by heavy quark symmetry the Isgur-Wise function is  
to leading order independent of both the 
heavy quark spin and the Dirac structure of the current, I use scalar quarks $Q$ and the 
current $Q^\dagger Q$.  With the standard heavy quark velocity transfer variable 
$w-1 =P^2_{cm}/2m^2_Q=V^2_{cm}/2$, we then have as 
$m_Q \rightarrow \infty$
\begin {eqnarray}
\xi (w) &\equiv& (2\pi)^3 \langle P(+{\vec P_{cm} \over 2}) \vert Q^\dagger(0) Q(0) 
\vert  P(-{\vec P_{cm} \over 2})  \rangle \\
&=& \int d^3r \int d^2a_1 \cdot \cdot \cdot \int d^2a_N 
e^{-i\vec V_{cm} \cdot [(m_d+{1 \over 2}br)  \vec r + {br \over \pi } \sqrt{2 \over {N+1}}
\sum_p {1 \over p}\vec a_p]}  \nonumber \\
&&
\vert \psi_0^{(0)}(\vec r) \vert^2 
\vert \chi_{0_1}(a_1^1) \vert^2 
\vert \chi_{0_1}(a_1^2) \vert^2 
\cdot \cdot \cdot   
\vert \chi_{0_N}(a_N^1) \vert^2 
\vert \chi_{0_N}(a_N^2) \vert^2 ~~.
\label{eq:xi}
\end {eqnarray}
Since we are interested in calculating the slope $\rho ^2$ in 
\begin {equation}
\xi (w) \simeq 1 - \rho^2 (w-1) + \cdot \cdot \cdot 
\end {equation}
we can expand the exponential to order $V_{cm}^2$ to get
\begin {eqnarray}
\xi (w) & \simeq &1 -{1 \over 2} V_{cm}^i V_{cm}^j
 \int d^3r \int d^2a_1 \cdot \cdot \cdot \int d^2a_N 
[(m_d+{1 \over 2}br)^2  r^i r^j 
+ {2b^2r^2 \over {\pi^2 (N+1)}} 
\sum_{p \alpha} {(a_p^\alpha)^2 \over p^2}  \hat e^i_\alpha \hat e^j_\alpha] \nonumber \\
&&
\vert \psi_0^{(0)}(\vec r) \vert^2 
\vert \chi_{0_1}(a_1^1) \vert^2 
\vert \chi_{0_1}(a_1^2) \vert^2 
\cdot \cdot \cdot   
\vert \chi_{0_N}(a_N^1) \vert^2 
\vert \chi_{0_N}(a_N^2) \vert^2  
\label{eq:xi(w)} \\
&& \nonumber \\
&\simeq&1-{ {\langle (m_d+{1 \over 2}br)^2  r^2 \rangle} \over 3}(w-1) - \rho^2_{transverse}(w-1) \cdot \cdot \cdot 
\label{eq:xi}
\end {eqnarray}
where 
\begin {eqnarray}
\rho^2_{transverse}(w-1) & \equiv & {1 \over 2} V_{cm}^i V_{cm}^j
 \int d^3r \int d^2a_1 \cdot \cdot \cdot \int d^2a_N 
\Bigl[{2b^2r^2 \over {\pi^2 (N+1)}} 
\sum_{p \alpha} {(a_p^\alpha)^2 \over p^2} \hat e^i_\alpha \hat e^j_\alpha \Bigr] \nonumber \\
&& 
~~~~~~~~~\vert \psi_0^{(0)}(\vec r) \vert^2 
\vert \chi_{0_1}(a_1^1) \vert^2 
\vert \chi_{0_1}(a_1^2) \vert^2 
\cdot \cdot \cdot   
\vert \chi_{0_N}(a_N^1) \vert^2 
\vert \chi_{0_N}(a_N^2) \vert^2 
\end {eqnarray}
is the new contribution to $\rho^2$ from flux tube transverse oscillations.  (Note that Eq. (\ref{eq:xi})
indicates that
the ordinary longitudinal contribution $m^2_d  \langle r^2 \rangle / 3$ 
should be supplemented with a term
which includes the flux tube contribution to the center of mass.  While keeping ${1 \over 2} br$ in 
comparison with $m_d$ in this term is not obviously consistent with the $br / m_d$ expansion 
we used to justify the adiabatic treatment of transverse oscillations relative to  longitudinal motion,
I believe this {\it is} the next term in the expansion of the longitudinal motion of $\vec r_Q$.
I nevertheless refrain from interpreting the extra ${1 \over 2} br$ term as a ``new" contribution to 
charge radii since it seems likely that the constituent quark mass of quark models has been chosen to at least
include a part of its effect. Thus with $m_d$ interpreted as the light constituent  quark
mass, Eq. (\ref{eq:r2correction}) rather than (\ref{eq:xi}) should be used.)  
With Eqs. (\ref{eq:transvectors}), the required angular averages may be done to obtain
\begin {equation}
\rho^2_{transverse}={{2b \langle  r^2 \rangle} \over 3 \pi^3} \zeta(3)
\label{eq:answer}
\end {equation}
as claimed in Eq.(\ref{eq:rho2correction}).  The previously quoted result 
(\ref{eq:r2correction}) for the transverse correction to a normal charge radius
follows by exactly parallel steps to those just outlined but  avoiding the use of the approximation
$\vec P_{cm} \simeq m_Q \vec V_{cm}$.  

    Both (\ref{eq:r2correction}) and (\ref{eq:rho2correction})  are simply the results of averaging over
the zero-point fluctuations of the $\vec a_p$, and could perhaps be obtained more simply if 
$\rho^2_{transverse}$ and $r^2_{transverse}$  were our only interests.  However, with this 
machinery, we can do much more.  For example, Eq. (\ref{eq:xi(w)}) gives us the complete function $\xi(w)$.  The 
new flux tube contribution to this function represents the probability that, in suffering the action of the current
$Q^\dagger Q$, the gluonic degrees of freedom were {\it not} excited.  At large recoils, probing short
distances, these flux tube excitations will become gluon jet events of perturbative QCD, and the analogous
probability of {\it not} emitting gluon radiation will be the Sudakov form factor \cite{Sudakov}.  In heavy quark
systems this probability appears in the (scale-dependent) contribution to $\rho^2$
\begin {equation}
\Delta \rho^2_{pert} \simeq {16 \over 81}ln \Bigl[ {{\alpha_s(\mu)} \over {\alpha_s(m_c)}}\Bigr]
\end {equation}
which arises from the velocity-dependent anomalous dimension of the heavy quark effective theory
matching coefficients \cite{vdependentd,Sudakov}.  
If one assumes that $\mu$ can be taken as low as the quark model scale, then
$\Delta \rho^2_{pert} \simeq0.13 \pm 0.05$ \cite{ISGW2}, quite 
comparable to the strong QCD contribution to $\rho^2$ derived here.
While we will not display or further discuss the full function $\xi(w)$ here, the machinery we have introduced is also
required to explicitly compute the rates of excitation of the hybrid states that compensate for $\rho^2_{transverse}$
to satisfy Bjorken's sum rule.  We now turn to that task.

	The inelastic form factor for the production of the hybrid $H_{n \ell m}^{(1 \pm)}$ is
\begin {eqnarray}
\tau_{\pm n \ell m}(w) &\equiv & (2\pi)^3 \langle H_{n \ell m}^{(1 \pm)}(+{\vec P_{cm} \over 2}) \vert Q^\dagger(0) Q(0) 
\vert  P(-{\vec P_{cm} \over 2})  \rangle \\
&=& \int d^3r \int d^2a_1 \cdot \cdot \cdot \int d^2a_N 
e^{-i\vec V_{cm} \cdot [(m_d+{1 \over 2}br)  \vec r + {br \over \pi } \sqrt{2 \over {N+1}}
\sum_p {1 \over p}\vec a_p]}  \nonumber \\
&&
\psi_{n \ell m}^{(1 \pm)}(\vec r)^* \psi_0^{(0)}(\vec r) \alpha_1(a_1^1 \mp i a_1^2)
\vert \chi_{0_1}(a_1^1) \vert^2 
\vert \chi_{0_1}(a_1^2) \vert^2 \cdot \cdot \cdot   
\vert \chi_{0_N}(a_N^1) \vert^2 
\vert \chi_{0_N}(a_N^2) \vert^2 .
\end {eqnarray}
To order $(w-1)$ this integral is nonzero only through the term $\vec V_{cm} \cdot \vec a_1$ in the expansion of the 
exponential, so we have
\begin {eqnarray}
\tau_{\pm n \ell m}(w) &\simeq& -{ib \alpha_1 \over \pi } \sqrt{2 \over {N+1}}
\int d^3r \int d^2a_1 
[\vec V_{cm} \cdot \vec a_1]  \nonumber \\
&&
\psi_{n \ell m}^{(1 \pm)}(\vec r)^* \psi_0^{(0)}(\vec r) r (a_1^1 \mp i a_1^2)
\vert \chi_{0_1}(a_1^1) \vert^2 
\vert \chi_{0_1}(a_1^2) \vert^2 \\
&\simeq& -i \sqrt{b \over 2\pi^3}
\int d^3r 
\psi_{n \ell m}^{(1 \pm)}(\vec r)^* \psi_0^{(0)}(\vec r) r \nonumber \\
&&
\Bigl[
({{1+cos \theta}\over 2})V_{cm-}-sin \theta e^{-i \phi}V_{cm~z}-({{1-cos \theta}\over 2})e^{-2i \phi}V_{cm+}
\Bigr]
\end {eqnarray}
which implies that only $\ell=1$ states are needed to saturate Bjorken's
sum rule \cite{IWonBj}.  Using Eqs. (\ref{eq:psi11hybrid})-(\ref{eq:psi1-1hybrid}), it follows that
\begin {equation}
\tau_{\pm n \ell m}(w)=\tau_n(w) \delta_{\ell1} V_{cm~m}^*
\end {equation}
where $V_{cm~+}\equiv -(V_{cm~x}+iV_{cm~y})$, $V_{cm~0}\equiv \sqrt{2} V_{cm~z}$, 
and $V_{cm~-}\equiv -(V_{cm~x}-iV_{cm~y})$ and where
\begin {equation}
\tau_n(w) =-i\sqrt{b \over 6\pi^3 m_d^2} \int dr r^3 \tilde \psi_{H^{(1)}_n}( r)^* \tilde \psi_P( r)
\end {equation}
with $\tilde \psi_{H^{(1)}_n}(r)$ and $\tilde \psi_P(r)$ 
the pure radial parts of the wavefunctions $\psi_{n \ell m}^{(1 \pm)}(\vec r)$
and  $\psi_0^{(0)}(\vec r)$, respectively, orthonormalized 
in their sectors to $\int dr r^2 \tilde \psi_m( r)^* \tilde \psi_n( r)=\delta_{mn}$.  
The total rate of transitions
to both $\pm$ ``helicity" states and all magnetic substates in units of the elastic rate is therefore
\begin {equation}
R_{H^{(1)}_n}={4b \over 3\pi^3}(w-1)\vert \int dr r^3 \tilde \psi_{H^{(1)}_n}( r)^* \tilde \psi_P( r)  \vert^2~~.
\end {equation}
It follows that
\begin {equation}
\sum_n R_{H^{(1)}_n}={4b \langle r^2 \rangle \over 3\pi^3}(w-1)
\end {equation}
since the $\psi_{H^{(1)}_n}( r)$ are a complete set of orthonormal states over $r$.  
On the other hand, the {\it loss} of rate from the 
elastic channel due to transverse modes is (in units of the elastic rate) $2\rho^2_{transverse}$, 
so we see on examining
Eq. (\ref{eq:answer}) that the total rate into the $p=1$ modes exactly compensates the $p=1$ term in the loss of 
elastic rate.

	Note that the rate to the hybrids $H^{(1)}_n$ is not highly fragmented.  Using ``canonical" parameters \cite{IP},  it is
straightforward to calculate the ratio
\begin {equation}
r \equiv { {\vert \int dr r^3 \tilde \psi_{H^{(1)}_0}( r)^* \tilde \psi_P( r)  \vert^2} \over {\langle r^2 \rangle} }
\end {equation}
which determines the fractional rate to the lowest-lying $p=1$ hybrid mesons.  This ratio is about
$40 \%$, so a substantial fraction of the total hybrid production will go into the lightest states.
Unfortunately, these heavy-light states cannot have the $J^{PC}$ exotic quantum numbers which
create ``smoking gun" signals for exotics in the $u \bar u$, $u \bar d$, $d \bar d$, $s \bar s$,
$c \bar c$, ... sectors.  This fact will make their observation (and in particular distinguishing them
from normal $c \bar d$ resonances) difficult in $\bar B$ decay.  Moreover, their expected 3 GeV masses
will substantially reduce the phase space available to their semileptonic decay, suppressing their
production rate in $\bar B$ decay well below that expected in the heavy quark limit,
and thereby violating duality \cite{NIonDuality}. 

\medskip
\section {Generalizations and Conclusions}

\medskip
	In this paper I have focussed on heavy quark semileptonic decay where exact duality {\it via} Bjorken's
sum rule allowed the derivation of a precise relation between transverse contributions to the position $\vec r_Q$
of the heavy quark and the rate of hybrid meson production.  The most important result of this
examination was the observation that since
\begin {equation}
{{\langle r_{Q\perp}^2 \rangle} \over {\langle r_Q^2 \rangle}}={2b \over \pi^3m_d^2}\zeta(3)
\end {equation}
is a pure number of order unity in QCD ($m_d$ is the constituent quark mass and so is proportional
to $b^{1 / 2}$), hybrid meson production in heavy quark semileptonic decay is unsuppressed by any
small parameters.

	While the lack of $J^{PC}$ exotic signals and phase space suppression may make the search for these 
states difficult in semileptonic $\bar B$ decay, the generalization of these rigorously derived results should
provide comfort and guidance to experimentalists planning to search for hybrids in other processes.
Consider first the $ \bar B$ system itself.  The factorizing 
components \cite{factorization} of $ \bar B$ hadronic decays to
charm, {\it e.g.} $ \bar B \rightarrow \pi^-X_c$ and $\bar B \rightarrow \rho^-X_c$, should populate hybrid states $X_c$ 
in the same proportions as in semileptonic decay.
The analogous statement holds for the rare factorizing $ \bar B$ hadronic decays
induced by an underlying 
$b \rightarrow u$ transition, {\it e.g.},
$ \bar B \rightarrow \pi^-X_u$ and $ \bar B \rightarrow \rho^-X_u$, where in $B^-$ decay one would have the advantage
of producing $J^{PC}$ exotic hybrid mesons.  
(In contrast, nonfactorizing processes like $ \bar B \rightarrow \bar K X_{c \bar c}$
cannot be expected to copiously produce $c \bar c$ hybrids \cite{cbarchybrids} because the $c \bar c$ pair
is being created by a local effective operator.)
The next most immediate generalization is perhaps to
ordinary electroproduction.  In this process there is an analogue of Bjorken's sum rule \cite{CabibboRadicati} and thus
of all of our results.  In particular, $ep \rightarrow e'X$ should produce ordinary $N^*$'s and hybrid
baryons with comparable cross sections once the threshold for hybrid baryons is crossed.  By the
same token, meson electroproduction by electroexcitation of an exchanged meson (the Sullivan processes \cite{Sullivan})
should readily produce hybrid mesons.  
(Recall that for light quarks the hybrid fraction is four times larger than in the heavy
quark limit.) While more difficult to study experimentally, the advantage of meson electroproduction over baryon 
electroproduction is that the former process will produce $J^{PC}$ exotic signals while the latter cannot.

	I am convinced that there are also generalizations of the results of this paper much farther afield
than electroproduction.  Consider, for example, forward photoproduction.  In this process strong 
interactions convert the virtual vector meson component of the photon into excited mesons.  If
these interactions are dominantly with the quarks in the virtual vector mesons, they will induce
dynamical processes very similar to those described here.  If they are dominantly with the flux tube,
then they can directly excite one of its phonon modes, and one can expect such direct hybrid production to be even stronger
than quark-recoil-induced production.  A very similar generalization applies to forward meson 
hadroproduction.  The main difference between hadroproduction and photoproduction is that the
quark spin of unity in the latter process is more likely to lead to the desirable $J^{PC}$ exotic
hybrids \cite{IKP}.

	In summary, I have identified an important new transverse contribution to hadronic charge radii from
flux tube zero-point motion.  Since the transverse quark motion has the universal QCD scale
$b^{1 \over 2}$, its contribution to hadronic charge radii is comparable to 
that from longitudinal quark motion. There are consequently substantial
corrections to potential model estimates of charge radii.  Given that the transverse and longitudinal
radii are comparable, it follows by Bjorken's sum rule that hybrid production in the heavy quark 
limit of $Q_1 \rightarrow Q_2 \ell \bar \nu_{\ell}$
semileptonic decay will be comparable to ordinary resonance production.  I have explicitly discussed
the generalization of this result to a broad range of possible hybrid production processes, and argued
that hybrid searches underway and in planning can safely assume that their hybrid cross sections, unless
forbidden by general selection rules, 
will indeed be comparable to those of ordinary mesons with the same phase space factors.

\bigskip\bigskip\bigskip\bigskip

{

\end{document}